\documentclass[nofootinbib,twocolumn,preprintnumbers]{revtex4-1}

\pdfoutput=1
\usepackage{amsmath}
\usepackage{amssymb}
\usepackage{graphicx}
\usepackage{booktabs}
\usepackage{color}
\usepackage{bbold}
\usepackage{hyperref}
\usepackage{amsmath}
\usepackage{tabularx}
\usepackage{multirow}

\newcommand{\ptvec}{{\vec{p}_T}}
\newcommand{\pt}{p_T}

\newcommand{\noun}[1]{{\scshape #1}}

\newcommand{\minlo}{{\noun{MiNLO$^{\prime}$}}}
\newcommand{\minnlo}{{\noun{MiNNLO$_{\rm PS}$}}}

\newcommand{\stepone}{{Step\,I}}
\newcommand{\steptwo}{{Step\,II}}
\newcommand{\stepthree}{{Step\,III}}
\DeclareMathOperator{\Tr}{Tr}

\AtBeginDocument{
\heavyrulewidth=.08em
\lightrulewidth=.05em
\cmidrulewidth=.03em
\belowrulesep=.65ex
\belowbottomsep=0pt
\aboverulesep=.4ex
\abovetopsep=0pt
\cmidrulesep=\doublerulesep
\cmidrulekern=.5em
\defaultaddspace=.5em
}
\begin{document}

\title{Next-to-next-to-leading order event generation for top-quark
  pair production}

\preprint{MPP-2020-233 --- CERN-TH-2020-219 --- LAPTH-049/20}

\author{Javier Mazzitelli$^1$, Pier Francesco Monni$^2$, Paolo
  Nason$^3$, Emanuele Re$^4$, Marius Wiesemann$^1$, Giulia Zanderighi$^1$\vspace{1em}}

\affiliation{$^1$ Max-Planck-Institut f\"ur Physik, F\"ohringer Ring 6, 80805 M\"unchen, Germany}
\affiliation{$^2$ CERN, Theoretical Physics Department, CH-1211 Geneva 23, Switzerland}
\affiliation{$^3$ Universit\`a di Milano Bicocca and INFN, Sezione di
    Milano Bicocca, Piazza della Scienza 3, 20126 Milano, Italy}
\affiliation{$^4$ LAPTh, Universit\'e Grenoble Alpes, Universit\'e Savoie Mont Blanc, CNRS, 74940 Annecy, France}

\begin{abstract}
  The production of top-quark pairs in hadronic collisions is among
  the most important reactions in modern particle physics
  phenomenology and constitutes an instrumental avenue to study the
  properties of the heaviest quark observed in nature.
  The analysis of this process at the Large Hadron Collider relies
  heavily on Monte Carlo simulations of the final state events, whose
  accuracy is challenged by the outstanding precision of experimental
  measurements.
  In this letter we present the first matched computation of top-quark
  pair production at next-to-next-to-leading order in QCD with
  all-order radiative corrections as implemented via parton-shower
  simulations.
  Besides its intrinsic relevance for LHC phenomenology, this work
  also establishes an important step towards the simulation of other
  hadronic processes with colour charges in the final state.
\end{abstract}

\maketitle

Top quarks are the heaviest elementary particles observed in nature
and play a unique role in the Standard Model (SM) of particle physics.
The large Yukawa coupling of the top quark to the Higgs boson
establishes a special avenue in the exploration of the Higgs sector of
the SM~\cite{deFlorian:2016spz} and of new physics
signals~\cite{Brivio:2019ius}.  Moreover, the value of the top-quark
mass enters precision Electro-Weak (EW) tests~\cite{Baak:2014ora} and
theoretical considerations on the stability of our
universe~\cite{Degrassi:2012ry}.

At the Large Hadron Collider (LHC), top quarks are predominantly
produced by strong interactions in association with their own
anti-particle ($t\bar{t}$).  The large value of the top-quark mass is
such that its production dynamics is safely inside the region of
validity of QCD perturbation theory.  Remarkable theoretical
advancements in the past years have led to very accurate predictions
for this process. Specifically, fixed-order computations that rely on
a power expansion in the strong coupling constant $\alpha_s$ are known
up to next-to-next-to-leading order
(NNLO)~\cite{Baernreuther:2012ws,Czakon:2012zr,Czakon:2012pz,Czakon:2013goa,Czakon:2015owf,Czakon:2016ckf,Catani:2019iny,Catani:2019hip}
(also including the top-quark decays~\cite{Behring:2019iiv}), and
EW corrections have been computed up to next-to-leading
order (NLO)~\cite{Beenakker:1993yr,Bernreuther:2005is,Kuhn:2006vh,Denner:2016jyo,Czakon:2017wor}.
In specific kinematic regimes, a reliable perturbative description
requires the all-order resummation of large radiative
corrections~\cite{Zhu:2012ts,Li:2013mia,Catani:2014qha,Catani:2018mei,%
  Beneke:2011mq,Beneke:2012wb,Ju:2020otc}.  Some of the above
calculations have been consistently combined to obtain the
state-of-the-art predictions at the LHC~\cite{Czakon:2019txp}.  The
striking accuracy of experimental measurements of the top-quark mass
requires pushing theoretical calculations to the edge of what can be
achieved with perturbative methods (for recent reviews
see~\cite{Nason:2017cxd,Hoang:2020iah}), and motivates new studies of
non-perturbative aspects of top-quark physics (see
e.g.~\cite{Beneke:2016cbu,Hoang:2017btd,Hoang:2018zrp,FerrarioRavasio:2018ubr}).

The large number of top-quark pairs produced at the LHC has allowed
for very satisfactory tests of the theory, both for the inclusive
production cross section and for multi-differential
distributions~\cite{Aad:2020tmz,Aad:2020nsf,Aad:2019hzw,Aad:2019ntk,%
  Khachatryan:2016kzg,Sirunyan:2017uhy,
  Sirunyan:2018ucr,
  Sirunyan:2018ptc,
  Sirunyan:2017mzl}. 
These tests have paved the way to the exploitation of top
cross-section measurements for the extraction of fundamental
parameters of the Standard Model, such as $\alpha_s$
and Parton Density Functions (PDFs), the top mass itself and the top
Yukawa couplings (see
e.g.~\cite{Chatrchyan:2013haa,
  Klijnsma:2017eqp,
  Sirunyan:2019zvx,
  Aad:2019mkw,
  Sirunyan:2020eds,Sirunyan:2019nlw,
  Cooper-Sarkar:2020twv
}).

Experimental analyses involving $t\bar{t}$ production heavily rely
upon its fully exclusive simulation.  This is important not only for
the study of the production dynamics itself, but, due to the complex
final states that involve a combination of leptons, jets, $b$ hadrons
and missing energy, also for several SM processes and new physics
searches for which the $t\bar{t}$ process acts as background.  The
needed simulations rely on event generators, which combine a
prediction for the high-energy scattering, that produces the
$t\bar{t}$ pair, initial- and final-state QCD radiation at all
perturbative orders via parton shower (PS) algorithms and
hadronization models (for a review see~\cite{Buckley:2011ms}). These
event generators have been subject of considerable research, dealing
with the matching of NLO QCD calculations to PS and a consistent
description of the top-quark
resonance~\cite{Frederix:2012ps,Hoeche:2014qda,Cormier:2018tog,%
  Jezo:2015aia,Jezo:2016ujg,Frederix:2016rdc}.  Current research for
the improvement of event generators focuses upon the development of
more accurate
PS~\cite{Hoche:2017hno,Dulat:2018vuy,Dasgupta:2018nvj,Bewick:2019rbu,Dasgupta:2020fwr,Forshaw:2020wrq,Hamilton:2020rcu},
as well as a framework to combine NNLO computations with PS into a
consistent event generator (NNLO+PS in the following).

Different frameworks for NNLO+PS computations have been developed in
recent years in the context of colour-singlet
production~\cite{Hamilton:2012rf,Alioli:2013hqa,Hoeche:2014aia,Monni:2019whf,Monni:2020nks}.
However, nearly a decade after these developments, a NNLO+PS method to
deal with hadron-collider processes with colour charges in the final
state (e.g.\ $t\bar{t}$), which are considerably more complex, is
still missing.

In this letter, we show how a generator of the same type as the ones
developed for colour singlet production processes in
Refs.~\cite{Monni:2019whf,Monni:2020nks}, dubbed there \minnlo{}, can
be constructed for top-quark pair production.
Our work constitutes the first computation of this type for reactions
with coloured particles in the final state.


The {\sc MiNNLO$_{\rm PS}$} procedure involves three steps. The first
one (referred to as \stepone{} in the following) corresponds to the
generation of a $t \bar{t}$ pair plus one light parton (i.e. the
\textit{underlying Born} configuration) according to the {\sc POWHEG}
method
\cite{Nason:2004rx,Nason:2006hfa,Frixione:2007vw,Alioli:2010xd},
carried out at the NLO level, inclusive over the radiation of a second
light parton.

The second step (\steptwo{}) characterizes the {\sc MiNNLO$_{\rm PS}$}
procedure, and it concerns the limit in which the light partons in the
above calculation become unresolved (i.e. the underlying Born
degenerates into a $t \bar{t}$ configuration without light jets). In
this limit the calculation must be supplemented with an appropriate
Sudakov form factor and higher-order terms, so as to guarantee that
the simulation remains finite as well as NNLO accurate for inclusive
$t \bar{t}$ production.
Most of the novelties in this letter have to do with \steptwo{} and will be illustrated below.

In the third step (\stepthree{}), the kinematics of the second
radiated parton, accounted for inclusively in \stepone{}, is generated
according to the standard {\sc POWHEG} method, which guarantees that
the NLO accuracy of $t \bar{t}$+jet cross section is preserved.
From this point on, subsequent radiation is included by the parton
shower, with the constraint of having a transverse momentum softer
than that of the last {\sc POWHEG} real emission.


The starting point to achieve NNLO accuracy in \steptwo{} 
is the well known factorization theorem for $t\bar{t}$ pair production at small
transverse momentum $\pt\equiv|\ptvec |$ differential in the phase
space of the $t\bar{t}$ pair
$d\Phi_{\rm t\bar{t}}\equiv d {\bar x}_1 d {\bar x}_2 [d\Phi_2]$.
Here ${\bar x}_{1,2}=m_{t{\bar t}}/\sqrt{s} \,e^{\pm y_{t\bar t}}$,
with $y_{t\bar t}$ being the rapidity of the $t\bar t$ system,
$m_{t\bar{t}}$ its invariant mass, $[d\Phi_2]$ denotes the
Lorentz-invariant two-body phase space and $\sqrt{s}$ is the collider
centre-of-mass energy. It
reads~\cite{Zhu:2012ts,Li:2013mia,Catani:2014qha,Catani:2018mei}
\begin{align}
\label{eq:bspace}
\frac{d\sigma}{d^2\ptvec d \Phi_{\rm
  t\bar{t}}}&=\sum_{c=q,\bar{q},g}
  \frac{|M^{(0)}_{c\bar{c}}|^2}{2 m_{t\bar{t}}^2}\int\frac{d^2\vec{b}}{(2\pi)^2} e^{i \vec{b}\cdot
  \ptvec } e^{-S_c \left(\frac{b_0}{b}\right)}\notag\\
&\times\sum_{i,j}\Tr({\mathbf H}_c{\mathbf \Delta})\,
 \,({C}_{ci}\otimes f_i) \,({C}_{\bar{c} j}\otimes f_j) \,,
\end{align}
where $b_0=2\,e^{-\gamma_E}$, $b=|\vec{b}|$. $S_c$ is the Sudakov radiator
which also enters the description of the production of a colour
singlet system at small transverse momentum
\begin{equation}
  S_c\left(k\right) = \int_{k^2}^{m_{t\bar{t}}^2}\frac{dq^2}{q^2} \left[A(\alpha_s(q))\ln\frac{m_{t\bar{t}}^2}{q^2}+B(\alpha_s(q))\right]\,.
\end{equation}
The first sum in Eq.~\eqref{eq:bspace} runs over all possible flavour
configurations of the incoming partons $p_1$ of flavour $c$ and $p_2$
of flavour $\bar c$.
The collinear coefficient functions
$C_{ij}=C_{ij}(z,p_1,p_2,\vec{b};\alpha_s(b_0/b))$ describe the
structure of constant terms related to the emission of collinear
radiation, and the parton densities are denoted by $f_i$ and are
evaluated at $b_0/b$. The operation $\otimes$ denotes the standard
convolution over the momentum fraction $z$ carried by initial state
radiation.
The factor
$\Tr({\mathbf H}_c{\mathbf \Delta})\, \,({C}_{ci}\otimes f_i)
\,({C}_{\bar{c} j}\otimes f_j)$
has different expressions for the $q\bar q$ and $gg$ channels and has
here a symbolic meaning. In particular, it has a rich Lorentz
structure that we omit for simplicity in Eq.~\eqref{eq:bspace}, which
is a source of azimuthal correlations in the collinear
limit~\cite{Catani:2010pd,Catani:2014qha}.

All quantities in bold face denote operators in colour space, and the
trace $\Tr({\mathbf H}_c{\mathbf \Delta} )$ in Eq.~\eqref{eq:bspace}
runs over the colour indices. This term can be expressed conveniently
in the colour space formalism of Ref.~\cite{Catani:1996vz}, where the
infrared-subtracted amplitude $|M_{c\bar c}\rangle$ for the production
of the $t\bar{t}$ system is a vector in colour
space~\cite{Catani:2014qha,Catani:2019hip}. It reads
\begin{equation}
\label{eq:trace}
\Tr({\mathbf H}_c{\mathbf \Delta} ) = \frac{\langle M_{c\bar c}|{\mathbf \Delta}| M_{c\bar c}
  \rangle}{|M^{(0)}_{c\bar{c}}|^2}\,,
\end{equation}
where $|M^{(0)}_{c\bar{c}}|^2=\langle M _{c\bar{c}}^{(0)} |M
_{c\bar{c}}^{(0)} \rangle$.
The hard function ${\mathbf H}_c={\mathbf H}_c(\Phi_{\rm
  t\bar{t}};\alpha_s(m_{t\bar{t}}))$ is obtained from the subtracted
amplitudes and the ambiguity in its definition corresponds to using a
specific resummation scheme~\cite{Bozzi:2005wk}. We adopt here the
definition of Ref.~\cite{Catani:2014qha}.
The operator ${\mathbf \Delta}$ encodes the structure of the quantum
interference due to the exchange of soft radiation at large angle
between the initial and final state, and within the final state. It is
given by ${\mathbf \Delta}={\mathbf V}^\dagger{\mathbf D}{\mathbf V}$,
where~\cite{Catani:2014qha}
\begin{align}
\label{eq:soft}
{\mathbf V} &= {\cal
  P}\exp\left\{-\int_{b_0^2/b^2}^{m_{t\bar{t}}^2}\frac{dq^2}{q^2}{\mathbf
  \Gamma}_t(\Phi_{\rm t\bar{t}};\alpha_s(q))\right\}\,.
\end{align}
The symbol ${\cal P}$ denotes the path ordering (with increasing
scales from left to right) of the exponential matrix with respect to
the integration variable $q^2$. ${\mathbf \Gamma}_t$ is the anomalous
dimension accounting for the effect of real soft radiation at large
angles, and 
${\mathbf D}={\mathbf D}(\Phi_{\rm t\bar{t}},\vec{b};\alpha_s(b_0/b))$
encodes the azimuthal dependence of the corresponding constant terms,
and is defined such that $[{\mathbf D}]_\phi={\mathbb 1}$, where
$[\cdots]_\phi$ denotes the average over the azimuthal angle $\phi$ of
$\ptvec$.

All of the above quantities admit a perturbative expansion in a series
in $\alpha_s(\mu)/(2\pi)$, where $\mu$ is the scale indicated
explicitly in the argument of each function. We generically denote
these expansions as
\begin{equation}
\label{eq:series}
F(\{x\} ;\alpha_s(\mu))=
\sum_{i}\frac{\alpha_s^i(\mu)}{(2\pi)^i} F^{(i)}(\{x\})\,,
\end{equation}
with $\{x\}$ being any other set of arguments of
$F\equiv\{A,\,B,\,C_{ij},\,{\mathbf H}_c,\,{\mathbf D},\,{\mathbf
  \Gamma}_t,|M_{c{\bar c}}\rangle \}$.
We do not indicate explicitly the scale of the amplitude
$|M_{c{\bar c}}\rangle$. In this case, the expansion~\eqref{eq:series}
is in powers of $\alpha_s(m_{t{\bar t}})$, and each of its
perturbative coefficients $|M_{c{\bar c}}^{(i)}\rangle$ includes an
extra single power of $\alpha_s(\mu_R^{(0)})$ with
$\mu_R^{(0)} \sim m_{t {\bar t}}$.
The above coefficients $F^{(i)}$ up to two loops are given in
Refs.~\cite{Czakon:2008zk,Catani:2010pd,Catani:2011kr,Czakon:2011xx,Catani:2012qa,Gehrmann:2012ze,Baernreuther:2013caa,Li:2013mia,Catani:2014qha,Gehrmann:2014yya,Echevarria:2016scs,Angeles-Martinez:2018mqh,Luo:2019bmw,Catani:2019hip,Luo:2019hmp,SoftFunction}.
The two-loop coefficient ${\mathbf D}^{(2)}$ is irrelevant at NNLO for
observables averaged over $\phi$, since
$[{\mathbf D}]_\phi={\mathbb 1}$, and therefore we do not include it
here.

Expanding the second order term in the exponential of
Eq.~\eqref{eq:soft}, one can write
\begin{align}
\label{eq:soft_recast}
{\mathbf V} = {\cal
 P}\Bigg[&\exp\left\{-\int_{b_0^2/b^2}^{m_{t\bar{t}}^2}\frac{dq^2}{q^2}\frac{\alpha_s(q)}{2\pi}{\mathbf
   \Gamma}^{(1)}_t\right\}\\
&\!\!\!\! \times\left(1-\int_{b_0^2/b^2}^{m_{t\bar{t}}^2}\frac{dq^2}{q^2}\frac{\alpha^2_s(q)}{(2\pi)^2}{\mathbf
 \Gamma}^{(2)}_t \right)\Bigg]
  + {\cal O}({\rm N}^3{\rm LL})\notag\,,
\end{align}
where we neglected N$^3$LL corrections which do not contribute at NNLO
in Eq.~\eqref{eq:bspace}.
In the following we will denote by ${\mathbf V}_{\rm NLL}$ the right hand side of
Eq.~\eqref{eq:soft_recast} with ${\mathbf
 \Gamma}^{(2)}_t\rightarrow 0$.
This enters the description of the $\pt$ spectrum at NLL accuracy.

To make contact with the procedure described in
Ref.~\cite{Monni:2019whf} we need to simplify further the structure of
the term ${\mathbf H}_c{\mathbf \Delta}$, which encodes the difference
with the colour singlet case. Our goal is to obtain a closed formula
in $\pt$ space that retains NNLO accuracy.
To this order, we observe that we can take the
${\mathbf \Gamma}^{(2)}_t$ term in Eq.~\eqref{eq:soft_recast} out of
the path ordering symbol.  We then perform a rotation in colour space
to diagonalize ${\mathbf \Gamma}^{(1)}_t$ and evaluate the exponential
matrix in Eq.~\eqref{eq:soft_recast}. Eq.~\eqref{eq:bspace} can be
reorganized using
\begin{align}
 e^{-S_c \left(\frac{b_0}{b}\right)}& \Tr({\mathbf H}_c{\mathbf \Delta})=e^{-\hat{S}_c \left(\frac{b_0}{b}\right)}\frac{\langle
    M_{c\bar{c}}^{(0)} | \left({\mathbf V}_{\rm NLL}\right)^\dagger{\mathbf V}_{\rm NLL} |M_{c\bar{c}}^{(0)}
  \rangle}{|M^{(0)}_{c\bar{c}}|^2}\notag\\
&\times \Tr({\mathbf H}_c{\mathbf D}) + E(\Phi_{\rm t\bar{t}},\vec{b})+{\cal O}(\alpha_s^5)\,,
\label{eq:approx}
\end{align}
where the trace is to be interpreted as in Eq.~\eqref{eq:trace}. 
The Sudakov radiator $\hat{S}_{c}$ is obtained from $S_c$ via the
replacement
\begin{align}
  B^{(2)}&\to B^{(2)}
  +\,\frac{\langle
    M_{c\bar{c}}^{(0)} | {\mathbf \Gamma}^{(2)\,\dagger}_t +
  {\mathbf \Gamma}^{(2)}_t|M_{c\bar{c}}^{(0)}
  \rangle}{|M^{(0)}_{c\bar{c}}|^2}\\
& + \frac{1}{|M^{(0)}_{c\bar{c}}|^2}\,2\,\Re\left[\langle
    M_{c\bar{c}}^{(1)} | {\mathbf \Gamma}^{(1)\,\dagger}_t +
  {\mathbf \Gamma}^{(1)}_t|M_{c\bar{c}}^{(0)}
  \rangle\right]\notag\\
& - 2\frac{\langle
    M_{c\bar{c}}^{(0)} | {\mathbf \Gamma}^{(1)\,\dagger}_t +
  {\mathbf \Gamma}^{(1)}_t|M_{c\bar{c}}^{(0)}
  \rangle}{|M^{(0)}_{c\bar{c}}|^4}\Re\left[\langle
    M_{c\bar{c}}^{(1)} |M_{c\bar{c}}^{(0)}
  \rangle\right] \notag\,.
\end{align}
The remainder term $E(\Phi_{\rm t\bar{t}},\vec{b})$ in Eq.~\eqref{eq:approx} contributes at
order $\alpha_s^2\ln(m_{t{\bar t}} \,b)$, but it is irrelevant for our
computation since it vanishes upon azimuthal integration
(i.e. $[E]_{\phi}=0$). For this reason, we will ignore it in the
following.
We then obtain
\begin{align}
\label{eq:starting}
\frac{d\sigma}{d^2\ptvec d \Phi_{\rm
  t\bar{t}}}&=\frac{1}{2 m_{t\bar{t}}^2}\sum_{c=q,\bar{q},g}
  \int\frac{d^2\vec{b}}{(2\pi)^2} e^{i \vec{b}\cdot
  \ptvec } e^{-\hat{S}_c \left(\frac{b_0}{b}\right)}\\
&\!\!\!\times \langle
    M_{c\bar{c}}^{(0)} | \left({\mathbf V}_{\rm NLL}\right)^\dagger{\mathbf V}_{\rm NLL} |M_{c\bar{c}}^{(0)}
  \rangle\notag\\
&\!\!\!\times\sum_{i,j}\Tr({\mathbf H}_c{\mathbf D})\, \,({C}_{ci}\otimes f_i)
  \,({C}_{\bar{c} j}\otimes f_j)+{\cal O}(\alpha_s^5)\notag\,.
\end{align}
This expression has almost the structure needed in order to carry out
the same procedure used in the colour singlet
case~\cite{Monni:2019whf,Monni:2020nks}, except for the ${\mathbf H}_c$
function which needs to be evaluated at the scale $b_0/b$ rather than
$m_{t{\bar t}}$. To the relevant accuracy we can perform this scale
change provided $B^{(2)}$ in $\hat{S}_c$ is also modified as
follows~\cite{Hamilton:2012rf,Monni:2019whf} (see e.g. Eq.~(4.25) of
Ref.~\cite{Monni:2019whf})
\begin{equation}
\label{eq:finalB2}
 B^{(2)} \to B^{(2)}
         +  2\pi\beta_0\,\frac{2\,\Re\left[\langle
    M_{c\bar{c}}^{(1)} |M_{c\bar{c}}^{(0)}\rangle\right]}{|M^{(0)}_{c\bar{c}}|^2} 
  \,.
\end{equation}
In the colour basis in which ${\mathbf \Gamma}^{(1)}_t$ is diagonal, the
matrix element
$\langle M_{c\bar{c}}^{(0)} | \left({\mathbf V}_{\rm
    NLL}\right)^\dagger{\mathbf V}_{\rm NLL} |M_{c\bar{c}}^{(0)}
\rangle$
is a linear combination of complex exponential terms, each of which
has the same factorized structure used as a starting point in the
appendix of Ref.~\cite{Monni:2019whf}.

Using this observation, we finally integrate over $\vec{b}$ by
expanding the integrand about $b_0/b\sim \pt$. Noticing that the
matrix element
$\langle M_{c\bar{c}}^{(0)} | \left({\mathbf V}_{\rm
    NLL}\right)^\dagger{\mathbf V}_{\rm NLL} |M_{c\bar{c}}^{(0)}
\rangle$
does not depend on the azimuthal angle $\phi$, up to terms of
${\cal O}(\alpha_s^5)$ we can express the result as a total
derivative, leading to the final $\pt$ space formula
\begin{widetext}
\begin{align}
\label{eq:master}
  \frac{d\sigma}{d \pt\,d \Phi_{\rm t\bar{t}}}\! =\! \frac{d}{d
  \pt}\bigg\{\sum_{c}\,\frac{e^{-\tilde{S}_{c}(\pt)}}{2 m_{t\bar{t}}^2}\,\langle
  M_{c\bar{c}}^{(0)} | \left({\mathbf V}_{\rm NLL}\right)^\dagger{\mathbf V}_{\rm NLL} |M_{c\bar{c}}^{(0)}
  \rangle\,\sum_{i,j}\left[\Tr(\tilde{\mathbf H}_c{\mathbf D})\, \,(\tilde{C}_{ci}\otimes f_i)
  \,(\tilde{C}_{\bar{c} j}\otimes f_j)\right]_\phi \bigg\}
  + R_{f} + {\cal O}(\alpha_s^5)\,.
\end{align}
\end{widetext}
To obtain Eq.~\eqref{eq:master}, we introduced the quantities
$\tilde{S}_c$, $\tilde{\mathbf H}_c$ and $\tilde{C}_{ij}$, which are
obtained by applying the transformations given in Eq.~(4.24) of
Ref.~\cite{Monni:2019whf} to $\hat{S}_c$, ${\mathbf H}_c$ and
$C_{ij}$. The latter quantities are now evaluated at the scale $\pt$.
The azimuthally averaged term $[\cdots]_\phi$ in
Eq.~\eqref{eq:master} is taken from the NNLO computation of the
$t{\bar t}$ cross section of
Refs.~\cite{Catani:2019iny,Catani:2019hip,SoftFunction} (see also
Ref.~\cite{hayk} for more details).
We also included the remainder $R_f=R_f(\pt)$, which denotes the
regular contribution to the $\pt$ distribution through
${\cal O}(\alpha_s^4)$, such that $\pt\,R_f(\pt)$ vanishes in the
$\pt\to 0$ limit.
The integral of Eq.~\eqref{eq:master} over $\pt$ provides a NNLO
accurate description of $t\bar t$ production differential in
$\Phi_{\rm t\bar{t}}$.

In order to build a Monte Carlo
algorithm for the generation of events with NNLO accuracy,
we have to modify the formula for the underlying born cross section of \stepone{}
in such a way that it matches 
Eq.~\eqref{eq:master} maintaining its NNLO accuracy. 
%
The procedure is described in detail in
Refs.~\cite{Hamilton:2012np,Hamilton:2012rf,Monni:2019whf,Monni:2020nks}
and for this reason we omit it here. 
%
For the practical implementation, we use the NLO+PS code for
$t\bar{t}+$jet of Ref.~\cite{Alioli:2011as}, and apply the \minnlo{}
procedure for heavy-quark pair production given in this letter. The PS
simulation is obtained with {\sc Pythia 8}~\cite{Sjostrand:2004ef},
without the modelling of non-perturbative effects, and under the
assumption of stable top quarks.
We stress that up to Eq.~\eqref{eq:finalB2} we retained also NLL
accuracy in the $\pt$ spectrum, while Eq.~\eqref{eq:master} is
strictly LL accurate.
Higher logarithmic accuracy could in principle be maintained in
Eq.~\eqref{eq:master}. However, this higher accuracy would be spoiled
by the PS used here, which is limited to LL.
On the other hand, Eq.~\eqref{eq:master} also preserves the class of
NLL corrections associated with the coefficient $A^{(2)}$ in the
Sudakov, that are traditionally included in PS
algorithms~\cite{Catani:1990rr}.
The formulation of a (N)NLO matching to PS that preserves logarithmic
accuracy beyond LL is still an open problem.

In the phenomenological study presented below, we consider LHC
collisions with a center of mass energy of $13$\,TeV.
The top-quark pole mass is set to $173.3$\,GeV and we consider five
massless quark flavours using the corresponding NNLO set of the
NNPDF31 \cite{Ball:2017nwa} parton densities with
$\alpha_s(m_Z)=0.118$.
The renormalization scale for the two powers of the strong coupling
constant entering the Born cross section is set to
$\mu_R^{(0)}= K_R \,m_{t\bar{t}}/2$.
In the rest of Eq.~\eqref{eq:master}, we implement the renormalization
($\mu_R=K_R\,\mu_0$) and factorization ($\mu_F=K_F\,\mu_0$) scale
dependence as described in Ref.~\cite{Monni:2019whf}, with the central
scale $\mu_0=m_{t\bar{t}}/2\, e^{-L}$ (hence replacing the scales set
to $\pt$ in Eq.~\eqref{eq:master}), where we defined $L=\ln Q/\pt$ and
$Q=m_{t\bar{t}}/2$.
The logarithm $L$ is turned off in the hard region of the $\pt$
spectrum so that 
the total derivative in Eq.~\eqref{eq:master} smoothly vanishes for
$\pt\gtrsim Q$ as in
Refs.~\cite{Bozzi:2005wk,Banfi:2012yh,Banfi:2012jm,Monni:2016ktx,Bizon:2017rah}. The
dependence of Eq.~\eqref{eq:master} on $\mu_R$, $\mu_F$ and $Q$ is of
order ${\cal O}(\alpha_s^5)$.
At small $\pt$ the scale of the strong coupling and the parton
densities is smoothly frozen around $Q_0=2$ GeV following the
procedure of Ref.~\cite{Monni:2020nks} to avoid the Landau
singularity.
To estimate the scale uncertainties we vary $K_R$ and $K_F$ by a
factor of 2 around their central value, while keeping
$1/2\le K_R/K_F\le 2$.
Results with a different central scale choice are reported in Ref.~\cite{supplmat}.

For comparison, we consider results from the fixed-order NNLO
calculation of Ref.~\cite{Catani:2019hip,Catani:2019iny} obtained with
the {\sc Matrix} code~\cite{Grazzini:2017mhc} using
$\mu_0= m_{t\bar{t}}/2$. Furthermore, we also show \minlo{} results,
obtained with the NLO+PS generator for $t\bar{t}$ plus zero and one
jet, constructed by turning off the NNLO corrections in
Eq.~\eqref{eq:master}.
The latter constitutes a new calculation as well.
\begin{table}[htp!]
  \vspace*{0.3ex}
  \begin{center}
\begin{tabular}{ccc}
\toprule
\minlo{} & NNLO & \minnlo{} \\
\midrule
$\;\,695.6(3)_{-17\%}^{+22\%}$\,pb& $\;\,769.8(9)_{-6.5\%}^{+5.0\%}$\,pb& $\;\,775.5(2)_{-7.2\%}^{+9.8\%}$\,pb\\
\bottomrule
\end{tabular}
\end{center}
\vspace{-1em}
  \caption{
    The total $t\bar{t}$ cross section in different approximations. 
    The quoted errors represent the
    scale uncertainty, while the numbers in brackets are the 
    numerical uncertainty on the last digit.}
\label{tab:XS}
\end{table}

Table~\ref{tab:XS} shows the total cross section for top-quark pair
production for \minlo{}, NNLO and \minnlo{}. The central \minlo{}
result is about $10.3$\% ($9.6$\%) smaller than the \minnlo{} (NNLO)
prediction and features much larger scale scale uncertainties. The
\minnlo{} result instead agrees with NNLO at the sub-percent level,
well within the perturbative uncertainties.  Small numerical
differences are expected even for inclusive observables, since the
\minnlo{} and NNLO calculations differ by terms beyond accuracy.

 \begin{figure*}[t]
     \centering
     \begin{tabular}{ccccc}
   \includegraphics[width=0.32\textwidth,page=6]{all_plots_paper.pdf} &&%
   \includegraphics[width=0.32\textwidth,page=4]{all_plots_paper.pdf} &&%
   \includegraphics[width=0.32\textwidth,page=5]{all_plots_paper.pdf}\\%
   \includegraphics[width=0.32\textwidth,page=2]{all_plots_paper.pdf} &&%
   \includegraphics[width=0.32\textwidth,page=1]{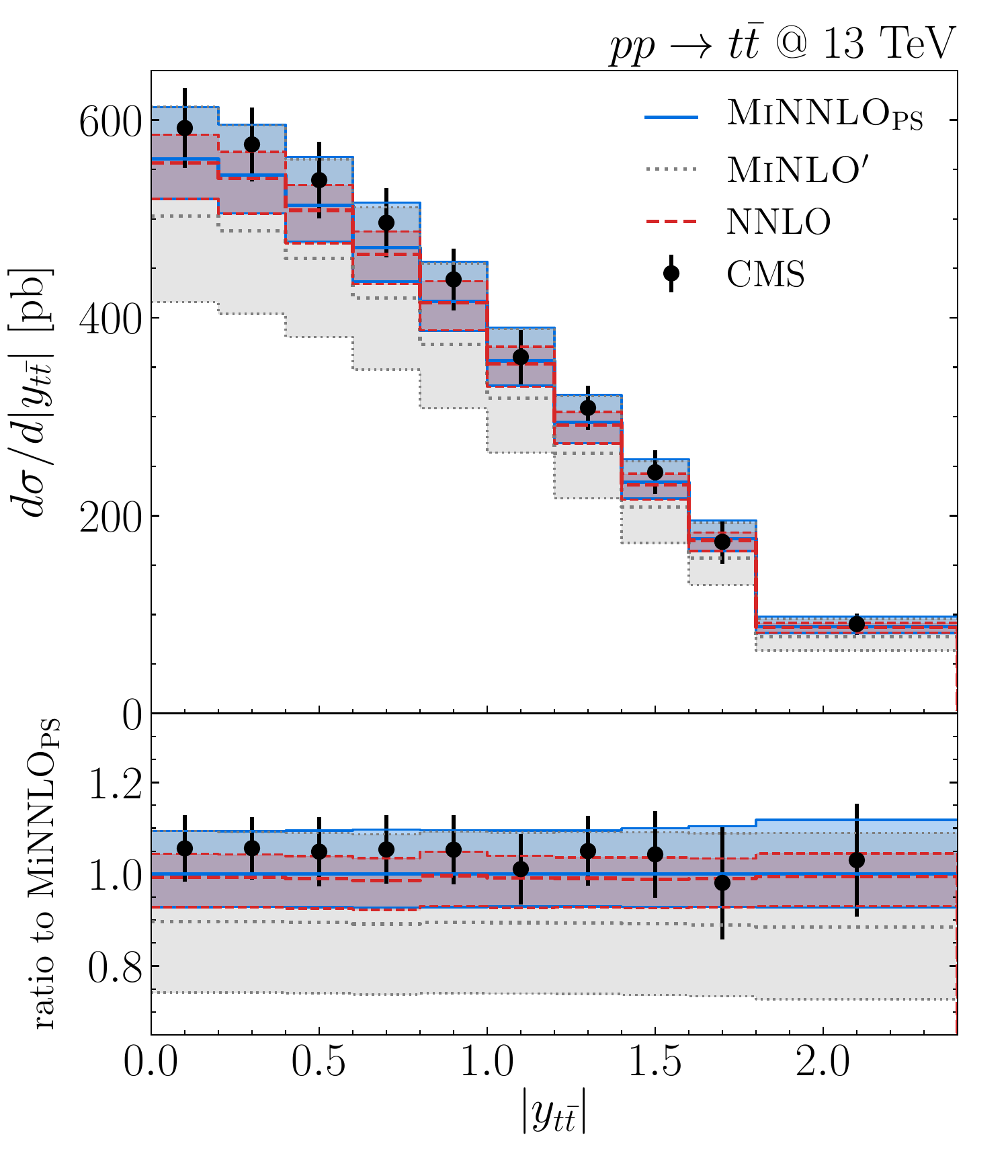} &&%
   \includegraphics[width=0.32\textwidth,page=3]{all_plots_paper.pdf}\\%
      \end{tabular}
      \caption{ \label{fig:dist} Distribution in the rapidity
        difference between the $t\bar{t}$ pair and the leading jet
        ($\Delta y_{t{\bar t},j_1}$), in the rapidity
        ($y_{t_{\rm av}}$) and the average transverse-momentum
        ($p_{T,t_{\rm av}}$) of the top and the anti-top, as well as
        in the rapidity ($y_{t\bar{t}}$), in the invariant mass
        ($m_{t\bar{t}}$) and in the transverse momentum
        ($p_{T,t\bar{t}}$) of the $t\bar{t}$ system. Predictions are
        shown for \minnlo{} (blue, solid), \minlo{} (black, dashed)
        and at NNLO (red, dashed). The black data points represent the
        CMS measurement at $13$\,TeV of Ref.~\cite{Sirunyan:2018wem},
        where the $y_{t_{\rm av}}$ and $p_{T,t_{\rm av}}$
        distributions have been obtained with leptonically decaying
        top quarks.}
 \end{figure*}

 In Fig.~\ref{fig:dist} we examine a set of differential
 distributions. To validate \minnlo{}, we compare it to the NNLO
 prediction without fiducial cuts, which could lead to significant
 differences due to the PS. Experimental data from the CMS
 collaboration unfolded and extrapolated to the inclusive phase
 space~\cite{Sirunyan:2018wem}, and divided by the appropriate
 branching fractions, are also shown.
 The top--left plot shows the rapidity difference between the
 $t{\bar t}$ system and the leading jet defined with
 $p_{T,\, j_1}\geq 120\,$GeV. Both \minlo{} and \minnlo{} are formally
 NLO accurate in this case, and the agreement between them indicates
 that this accuracy is retained by the \minnlo{} procedure. The same
 conclusion holds for other observables that require at least one
 resolved hard jet.

 The distributions in the average top-quark rapidity
 ($y_{t_{\rm av}}$) and transverse momentum ($p_{T,t_{\rm av}}$) as
 well as in the invariant mass ($m_{t{\bar t}}$) and rapidity
 ($y_{t{\bar t}}$) of the $t{\bar t}$ pair shown in
 Fig.~\ref{fig:dist} are inclusive over QCD radiation. For such
 distributions \minnlo{} is expected to be NNLO accurate. Indeed,
 \minnlo{} and NNLO yield consistent results, with fully overlapping
 uncertainty bands. The small differences in the central value are
 once again due to the different treatment of terms beyond NNLO
 accuracy. The larger uncertainty bands of the \minnlo{}
 predictions are expected, since additional scale dependent terms are
 included within the first term in the r.h.s. of Eq.~\eqref{eq:master}
 that are not present in the fixed-order calculation. In comparison to
 the \minlo{} results the inclusion of NNLO corrections through
 \minnlo{} has an impact of about $10$\%--$20$\% on the differential
 distributions and substantially reduces the perturbative
 uncertainties. Also the agreement with data is quite remarkable. All
 data points are within one standard deviation from the \minnlo{}
 prediction, with the exception of the very first bin in the
 $m_{t{\bar t}}$ distribution that, on the other hand, is strongly
 affected by the finite width of the top, whose effects are not
 included here.

 We finally discuss the transverse-momentum spectrum of the $t\bar{t}$
 pair, denoted by $p_{T,\,{t\bar t}}$ in the bottom--right panel of
 Fig.~\ref{fig:dist}. At large transverse momenta, the three
 predictions considered are effectively NLO accurate. Indeed, \minlo{}
 and \minnlo{} are essentially indistinguishable in that region, and
 at the same time consistent with the spectrum at fixed order. The
 small differences with NNLO are due to the generation of further
 radiation by the PS.
 At small transverse momenta, \minnlo{} induces ${\cal O}(10\%)$
 corrections with respect to \minlo{} and significantly reduces the
 large scale dependence.  In this region, it also differs in shape
 from the NNLO calculation, which diverges and becomes unphysical for
 vanishing transverse momenta. Within the relatively large
 experimental errors, \minnlo{} slightly improves the description of
 the data in terms of shape compared to NNLO for this observable.

 In this letter we have presented the matching of the NNLO computation
 for top-quark pair production at hadron colliders with parton
 showers. This result has been obtained by constructing the \minnlo{}
 method for the production of heavy quarks, which constitutes the
 first NNLO+PS prediction for reactions with colour charges in the
 final state in hadronic collisions.
 The comparisons presented in Fig.~\ref{fig:dist} provide a numerical
 validation of \minnlo{} for top-quark pair production, demonstrating
 its NNLO accuracy. 
 The simulations presented here also allow for the inclusion of the
 top-quark decay, paving the way to an accurate event generation for
 $t{\bar t}$ production at the LHC which will enable precise
 comparisons of fiducial measurements to theory.
 \\[0.5cm]
 {\bf Acknowledgements.}  We would like to thank Stefano Catani,
 Massimiliano Grazzini, and Keith Hamilton for discussions and
 comments on the manuscript. P.N. acknowledges support from Fondazione
 Cariplo and Regione Lombardia, grant 2017-2070, and from INFN.

\bibliographystyle{apsrev4-1}
\bibliography{NNLO+PS_ttbar} 

\newpage

\onecolumngrid
\newpage
\appendix

\section*{Supplemental material}
\begin{figure*}[h]
     \centering
     \begin{tabular}{ccccc}
   \includegraphics[width=0.32\textwidth,page=6]{all_plots_paper_supplemental.pdf} &&%
   \includegraphics[width=0.32\textwidth,page=4]{all_plots_paper_supplemental.pdf} &&%
   \includegraphics[width=0.32\textwidth,page=5]{all_plots_paper_supplemental.pdf}\\%
   \includegraphics[width=0.32\textwidth,page=2]{all_plots_paper_supplemental.pdf} &&%
   \includegraphics[width=0.32\textwidth,page=1]{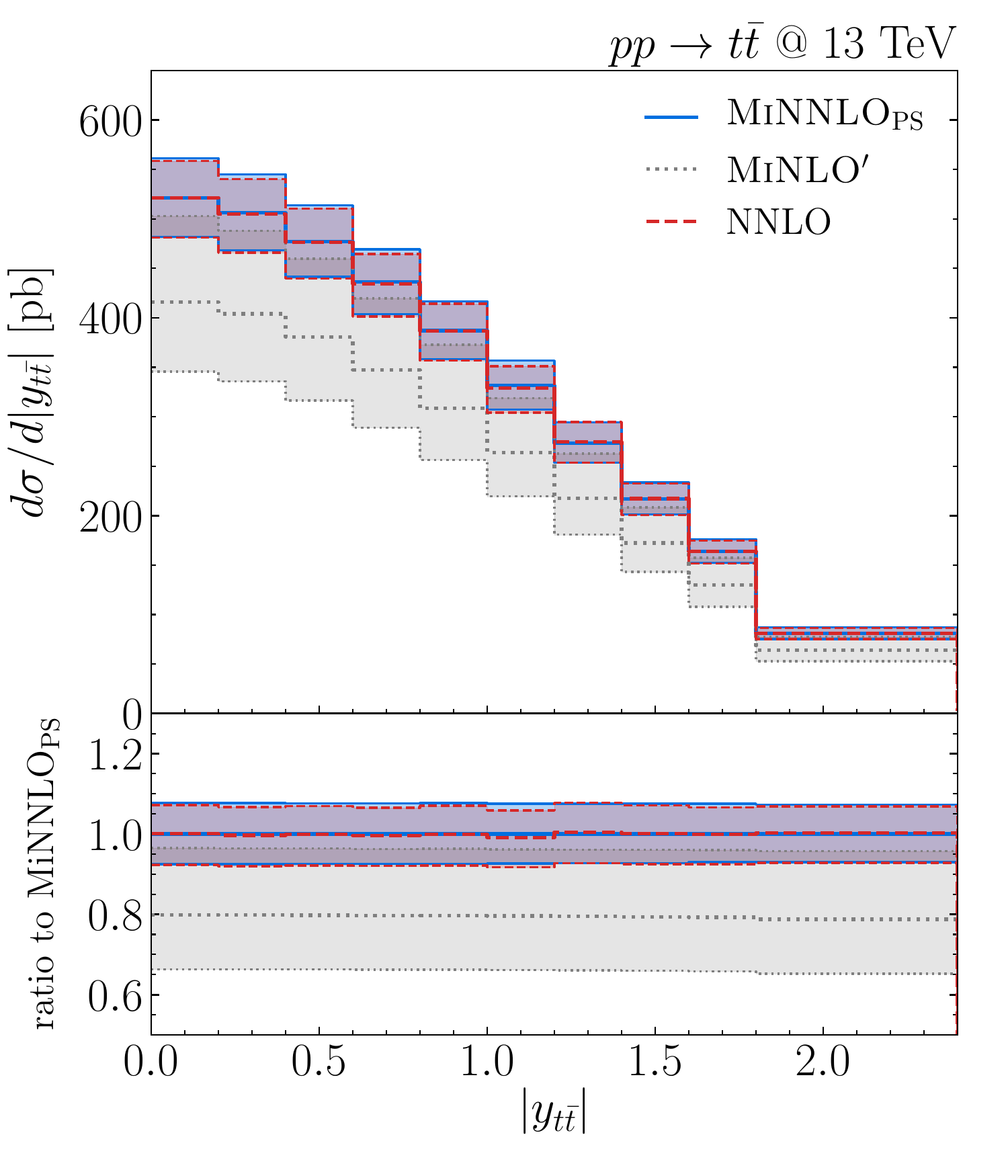} &&%
   \includegraphics[width=0.32\textwidth,page=3]{all_plots_paper_supplemental.pdf}\\%
      \end{tabular}
      \caption{ \label{fig:dist2} Same as Fig.\,1, albeit without data and with the scale setting $\mu_R^{(0)}= K_R \,m_{t\bar{t}}$ and $\mu_0 = m_{t\bar{t}}\,e^{-L}$ for \minnlo{} and \minlo{}, and $\mu_0 = m_{t\bar{t}}$
for NNLO.}
 \end{figure*}

In this appendix we complement the results presented in the letter by comparing \minnlo{}, \minlo{}, and NNLO predictions with different scale settings.
In particular, we set $\mu_R^{(0)}= K_R \,m_{t\bar{t}}$ and $\mu_0 = m_{t\bar{t}}\,e^{-L}$ for \minnlo{} and \minlo{}, and we use $\mu_0 = m_{t\bar{t}}$
for the NNLO fixed-order calculation. The other settings are as reported in the main text. Table~\ref{tab:XS2} reports the total cross sections, while
Fig.~\ref{fig:dist2} shows the same distributions as shown in the letter, but with the updated scale setting. Despite the fact that higher-order differences are 
expected for all observables, we observe an excellent agreement between \minnlo{} and NNLO predictions for this scale choice.
\begin{table}[htp!]
  \vspace*{0.3ex}
  \begin{center}
\begin{tabular}{ccc}
\toprule
\minlo{} & NNLO & \minnlo{} \\
\midrule
$\;\,572.9(2)_{-17\%}^{+21\%}$\,pb& $\;\,719.1(8)_{-7.6\%}^{+7.0\%}$\,pb& $\;\,719.8(2)_{-7.4\%}^{+7.6\%}$\,pb\\
\bottomrule
\end{tabular}
\end{center}
\vspace{-1em}
  \caption{
    The total $t\bar{t}$ cross section in different approximations. 
    The quoted errors represent the
    scale uncertainty, while the numbers in brackets are the 
    numerical uncertainty on the last digit.}
\label{tab:XS2}
\end{table}

\end{document}